\documentclass[amsmath,amssymb,amsfonts,aps,pre,superscriptaddress,showpacs,showkeys,longbibliography,10pt,twocolumn]{revtex4-1}

\usepackage{epsfig}
\usepackage{ulem}
\usepackage[english]{babel}
\usepackage{color}
\usepackage{hyperref}

\begin{document}
\title{Critical exponents of fluid-fluid interfacial tensions near a critical endpoint in a nonwetting gap}

\author{Joseph O. Indekeu}
\email[Author to whom correspondence should be addressed: ]{joseph.indekeu@kuleuven.be}
\affiliation{Institute for Theoretical Physics, KU Leuven, BE-3001 Leuven, Belgium}
\author{Kenichiro Koga}\email{koga@okayama-u.ac.jp}
\affiliation{Research Institute for Interdisciplinary Science, Okayama University, Okayama 700-8530, Japan}

\date{\today}


\begin{abstract}
 Fluid three-phase equilibria, with phases $\alpha, \beta, \gamma$, are studied close to a tricritical point, analytically and numerically, in a mean-field density-functional theory with two densities. Employing Griffiths' scaling for the densities, the interfacial tensions of the wet and nonwet interfaces are analysed. The mean-field critical exponent is obtained for the vanishing of the critical interfacial tension $\sigma_{\beta\gamma}$ as a function of the deviation of the noncritical interfacial tension $\sigma_{\alpha\gamma}$ from its limiting value at a critical endpoint $\sigma_{\alpha,\beta\gamma}$. In the wet regime, this exponent is $3/2$ as expected. In the nonwetting gap of the model, the exponent is again $3/2$, except for \textcolor{black}{the approach to} the critical endpoint on the neutral line where $\sigma_{\alpha\beta} = \sigma_{\alpha\gamma}$. \textcolor{black}{When this point is approached along any path with $\sigma_{\alpha\beta} \neq \sigma_{\alpha\gamma}$, or along the neutral line,} $\sigma_{\beta\gamma} \propto  |  \sigma_{\alpha\gamma} - \sigma_{\alpha,\beta\gamma}|^{3/4}$, featuring an anomalous critical exponent $3/4$, which is an exact result derived by analytic calculation and explained by geometrical arguments.
\end{abstract}

\maketitle

\section{Introduction}
Wetting and wetting phase transitions are topics of ever\-green significance in statistical mechanics and beyond (for reviews, see 
\cite{dG,SulTdG,Diet,FLN,BonnRoss,BonnRMP}). Recently there has been a revival of interest in predicting wetting phase diagrams from density-functional theory for adsorbed fluids \cite{PNAS}. Primarily two settings have been considered: fluid two-phase equilibria at a solid substrate \cite{PNAS,Parry1,Parry2}, and fluid three-phase equilibria (without a wall) in which the three phases are treated on an equal footing \cite{IK, Parry3, Leermakers, ITB}. In the latter setting, liquid mixtures have been studied near critical endpoints in the vicinity of a bulk tricritical point. Both experimentally \cite{KahlweitBusse,AratonoKahlweit,Erratum,KSB} and theoretically \cite{IK, Parry3} there is evidence for the existence of a nonwetting gap in which the standardly expected phenomenon of ``critical-point wetting" (CPW) does {\it not} take place. 

The behavior of interfacial tensions near bulk critical points, and its interplay with wetting, is a problem of long-standing interest \cite{Widom,LW,RW,Langetal,RW}. Cahn argued that for two phase-separated fluids adsorbed at a wall, one fluid phase must wet the interface between the other fluid phase and the wall, upon approach of a bulk critical point at which the two fluid phases become a single critical phase \cite{Cahn}. This CPW scenario has been confirmed in various systems of experimental and theoretical interest \cite{MC,Langetal,NF}. However, the argument in favor of its necessity has repeatedly been shown to be circular and counterexamples to CPW have gradually been receiving unbiased attention \cite{Pandit,NighInd,EbnerSaam,IndekeuCPD,Sevrin,KahlweitBusse,Ind}.

\section{Two-density Density-functional theory}
Various models of the structure and tension of interfaces at fluid three-phase equilibria were studied within mean-field density-functional theory (DFT) in \cite{KW}. Here, we concentrate on one of these models, akin to model T in \cite{KW}, analysed analytically and numerically in \cite{KI}, taking into account all the constraints on the densities imposed by the mean-field theory for tricritical points developed by Griffiths \cite{Grif}. The essential ingredients are recalled for clarity. 

For concreteness, a three-component fluid mixture is considered, for which the three chemical potentials $\mu_i$, $i \in {1,2,3}$, together with pressure $p$ and temperature $T$, span a 5-dimensional thermodynamic field space with a 4-dimensional manifold of equilibrium states. Therein, a 2-dimensional surface of three-phase coexistence of phases $\alpha$, $\beta$ and $\gamma$ is studied, bounded by two lines of critical endpoints (CEP) which meet at a tricritical point (TCP). At three-phase coexistence, $\alpha$, $\beta$ and $\gamma$ may meet at a common line of contact (nonwet state) or one of them, say $\beta$, may intrude between $\alpha$ and $\gamma$ (wet state) \cite{RW}. The transition between the nonwet and the wet states is the wetting phase transition.

The (dimensionless) interfacial free-energy density $\Psi$  of the DFT with two densities $\rho_1$ and $\rho_2$ is taken to be
\begin{equation}\label{Psi}
\Psi (\rho_1,\rho_2) = \frac{1}{2} \sum_{i=1,2}\left (\frac{d\rho_i}{dz}\right )^2   + \prod_{\nu=\alpha,\beta,\gamma}\, \sum_{i=1,2}(\rho_i - \rho_{i}^{\nu} )^2. 
\end{equation}
It is a product of three potential wells, centered on points in the $(\rho_1,\rho_2)$-plane, with bulk phase densities $\vec{\rho}_{\nu} \equiv (\rho_{1}^{\nu},\rho_{2}^{\nu})$, with $ \nu = \alpha, \beta,\gamma$. 
The equilibrium density profiles are those spatially varying densities $\rho_1 (z)$ and $\rho_2 (z)$ that minimize the interface free-energy functional $\int \, \Psi \, dz$ subject to two bulk boundary conditions, at $| z| \rightarrow \infty$. For example, for $\alpha$ at $z=-\infty$ and $\gamma$ at $z=\infty$,
\begin{equation}
    \sigma_{\alpha\gamma} = \min_{\rho_1 (z), \rho_2 (z)} \int_{-\infty}^{\infty}\Psi (\rho_1,\rho_2) \, dz.
 \end{equation}
If the $\alpha\gamma$ interface is wet by $\beta$, one has
\begin{equation}
\label{wet}
    \sigma_{\alpha\gamma} = \sigma_{\alpha\beta}+ \sigma_{\beta\gamma},
\end{equation}
whereas, if $\beta$ does not wet the $\alpha\gamma$ interface, one has
\begin{equation}
\label{nonwet}
    \sigma_{\alpha\gamma} < \sigma_{\alpha\beta}+ \sigma_{\beta\gamma}.
\end{equation}

A simple analytic expression was proposed and verified numerically for the interfacial tension of a nonwet interface in this model \cite{KI}. It reads
\begin{equation}
\label{conj}
    \sigma_{\alpha\gamma} = \frac{\sqrt{2}}{6} {\cal{P}}^3 \ell.
 \end{equation}
Here $\cal{P}$ is the length of the $\alpha\gamma$ edge in the three-phase $\alpha\beta\gamma$ triangle spanned by the bulk phase points in the $(\rho_1,\rho_2) $-plane, and $\ell$ the length of the median connecting $\beta$ to the $\alpha\gamma$ edge (see Figure 1 in \cite{KI}).
The conjecture \eqref{conj} was proven to be exact by mathematical solution of the model \cite{Parry3}.
 
In the Griffiths mean-field theory of tricritical phenomena, linear combinations of the densities, which we rename just $\rho_1$ (for the principal density) and $\rho_2$ (for the subsidiary density), obey a constraint on the bulk densities, in each phase, which takes the form
\begin{equation}
\label{constraint}
\rho^{\nu}_2= -(\rho^{\nu}_1)^2,
\end{equation} 
and the bulk principal densities (for the three phases in the coexistence region) are the solutions of the third-degree polynomial
\begin{equation}
\label{three}
    \phi(\rho_1) = \rho_1^3 - 3t\rho_1 + 2s.
\end{equation}
Here, $s$ and $t \,(>0)$ are linear combinations of the thermodynamic field variables $\mu_i$, $p$ and $T$, such that the TCP is at $s=t=0$ and the three-phase coexistence range for $s$ is
\begin{equation}
    - t^{3/2} < s < t^{3/2}.
\end{equation}
The two CEP lines in the $(s,t)$-plane satisfy $s t^{-3/2} = \pm 1$. They meet tangentially at the TCP in a cusp.
Varying $s$ between $-t^{3/2}$ and $t^{3/2}$ at constant $t$ can, for example, be thought of as varying pressure $p$ or varying one of the chemical potentials $\mu_i$ in order to interpolate between the two CEPs at constant temperature $T$.

\section{Interfacial tension in the three-phase region between critical endpoints}
For the critical interfacial tension close to a CEP at which $\beta $ and $\gamma$ become identical phases, i.e., close to $s = t^{3/2}$, an exact calculation was performed in \cite{KI}, which implies the following series expansion
\begin{eqnarray}
\label{criticalinterface}
    \sigma_{\beta\gamma} &=& \frac{16\sqrt{3}}{9}t^2(1+t)^{1/2}(1+4t)^{3/2} \epsilon^{3/2} \nonumber \\ &-& \frac{4\sqrt{3}\,t^2(1+4t)^{1/2}}{81(1+t)^{1/2}}(-1+55t+92t^2) \epsilon^{5/2}\nonumber \\ &+& {\cal O}(\epsilon^{7/2}),
\end{eqnarray}
where $\epsilon \,(\ll 1)$ is a measure of the distance to the CEP in the manner
\begin{equation}
\label{expand}
    s = (1-\epsilon) t^{3/2}.
\end{equation}
As noted in \cite{KI}, since $\epsilon$ is linear in the deviation of the field $s$ from its CEP value, one retrieves the standard mean-field critical exponent $3/2$ for the vanishing of the interfacial tension at two-phase criticality.

We now turn to the calculation of the general (non-critical) interfacial tension for a nonwet interface in the three-phase region of the 
$(s,t)$-plane. This can be obtained by direct solution, using \eqref{constraint} and \eqref{three}. Additionally, in order to gain more insight into the properties close to a CEP from analytic results, one may perform a series expansion in $\epsilon$. The critical exponent describing the vanishing of the deviation of the interfacial tension, say $\sigma_{\alpha\beta}$, of a nonwet interface from its value $\sigma_{\alpha,\beta\gamma}$ at the $\beta\gamma$ CEP was found to take the value 1 under general circumstances,
\begin{equation}
\label{deviation}
    \sigma_{\alpha\beta} - \sigma_{\alpha,\beta\gamma} \propto  \epsilon + {\cal O}(\epsilon^{3/2}).
\end{equation}
In the following this result is checked in the various regions of the wetting phase diagram uncovered in \cite{IK}. If \eqref{deviation} holds true, then together with \eqref{criticalinterface} it implies the generic relationship between the interfacial tensions of the nonwet interfaces,
\begin{equation}
\label{generic}
    \sigma_{\beta\gamma} \propto | \sigma_{\alpha\beta}-\sigma_{\alpha,\beta\gamma}|^{3/2}.
\end{equation}
This critical-endpoint exponent $3/2$ is expected in mean-field theory. 

\subsection{Wet states close to the TCP}
In this subsection, states in which the $\alpha\gamma$ interface is wet by $\beta$ are examined. According to the wetting phase diagram obtained in \cite{IK}, such states occur close to the TCP and a sufficient condition for wet states to span the entire range between the two CEPs at constant $t$ is $t < 2/\sqrt{3} -1 \approx  0.1547$. The states satisfy \eqref{wet}. Direct calculation confirms that \eqref{deviation} holds close to the CEP. 

\begin{figure}[h!]
\centering
\includegraphics[width=0.9\linewidth]{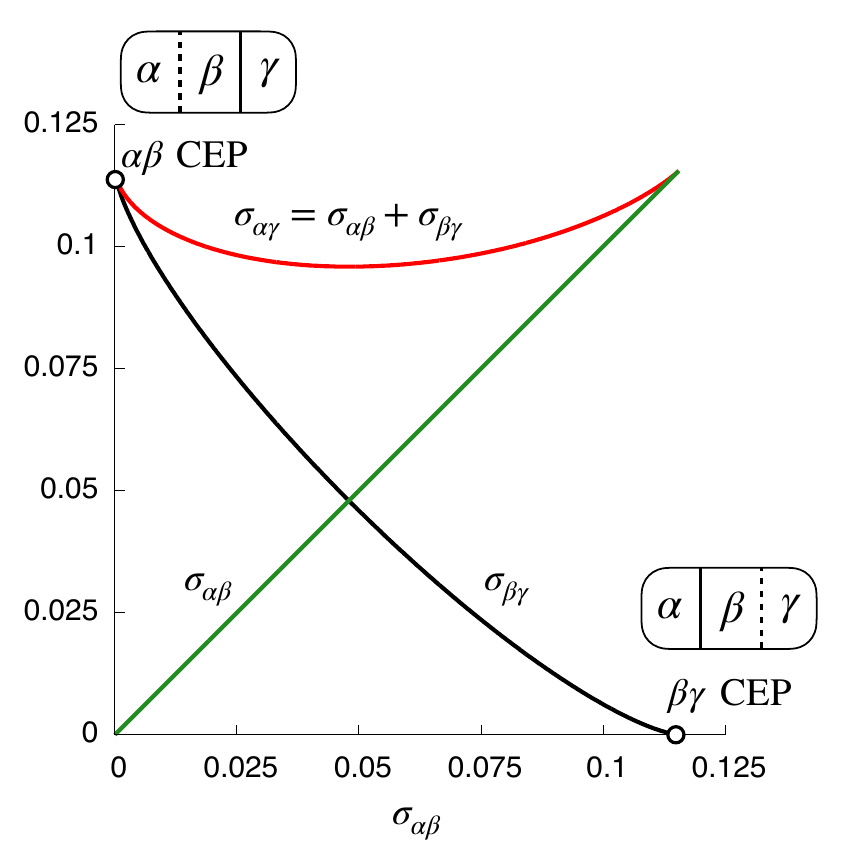}
\caption{Interfacial tensions $\sigma_{\alpha\beta}$ and $\sigma_{\beta\gamma}$ of the interfaces $\alpha\beta$ and $\beta\gamma$, respectively, as well as the interfacial tension $\sigma_{\alpha\gamma}$ of the $\alpha\gamma$ interface that is wet by $\beta$, versus $\sigma_{\alpha\beta}$ for a sweep (varying $s$) between the two CEPs at $t = 0.1$. In this regime close to the TCP, only wet states are encountered and CPW holds.  Note that the plot of $\sigma_{\beta\gamma}$ is convex towards the origin as was found experimentally in a specific fluid mixture \cite{Langetal}. The three-fluid interface configurations asymptotically close to the CEPs, on the edge of three-phase coexistence, are drawn in the diagrams. In these diagrams, the solid line represents the non-critical interface and the dashed line the diffuse near-critical interface.}
\label{fig:onetenth}
\end{figure}

For $t=0.1$ Fig.\ref{fig:onetenth} displays the interrelationship of the interfacial tensions $\sigma_{\alpha\beta}$ and $\sigma_{\beta\gamma}$. 
The generic law \eqref{generic} holds. It is also shown that the wetting condition \eqref{wet} is satisfied.

\subsection{Nonwet states inside the nonwetting gap}
The nonwetting gap, in which CPW is absent, extends in this model from $t = (7- \sqrt{3})/8 \approx 0.1569$ to $t= (7+ \sqrt{3})/8 \approx 1.5930$. As representative examples the results are shown of interpolating between the two CEPs at $t = 0.25$ and between the two CEPs at $t=1$. 

\begin{figure}[h!]
\centering
\includegraphics[width=0.9\linewidth]{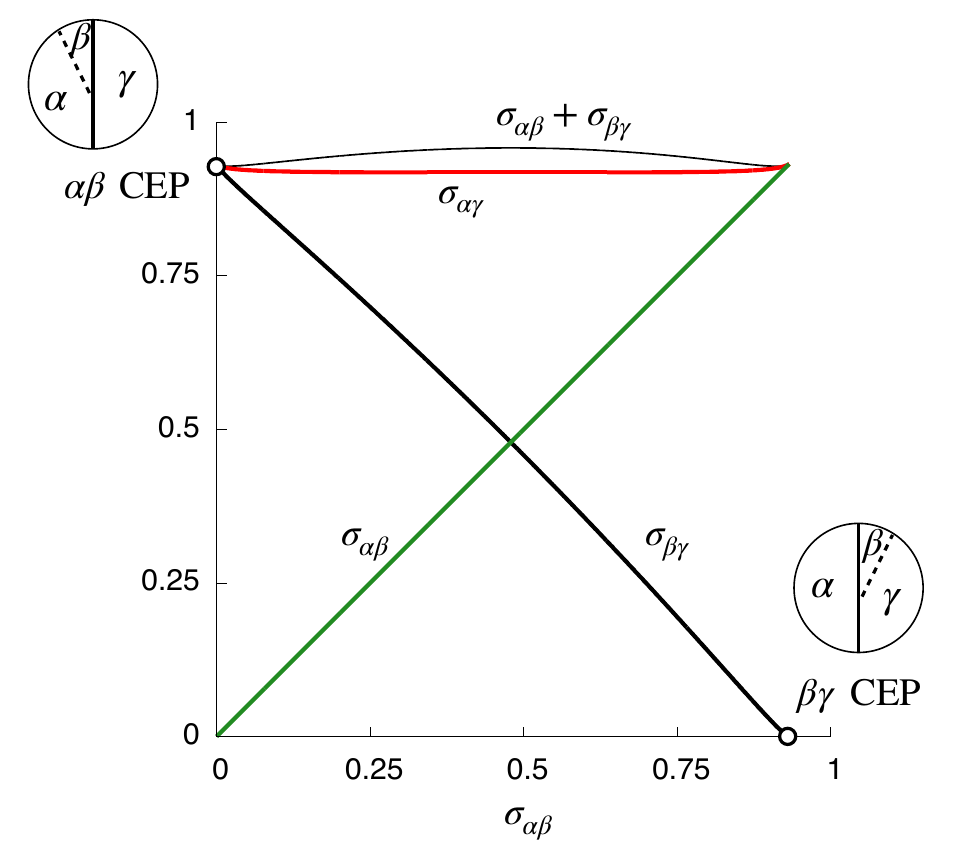}
\caption{Interfacial tensions $\sigma_{\alpha\beta}$, $\sigma_{\beta\gamma}$ and $\sigma_{\alpha\gamma}$ of the  interfaces $\alpha\beta$, $\beta\gamma$ and $\alpha\gamma$, respectively, versus $\sigma_{\alpha\beta}$ for a sweep (varying $s$) between the two CEPs at $t = 0.25$. Note that since $\sigma_{\alpha\gamma} < \sigma_{\alpha\beta} +\sigma_{\beta\gamma}$ only nonwetting states are encountered and CPW is absent. The CEPs lie in the nonwetting gap. Note that the plot of $\sigma_{\beta\gamma}$ is not convex towards the origin, and on this scale appears to be almost linear. However, asymptotically close to the CEPs $\sigma_{\beta\gamma}$ approaches the axes tangentially, and features the exponent 3/2 in accord with \eqref{generic}.  The three-fluid interface configurations asymptotically close to the CEPs, on the edge of three-phase coexistence, are drawn in the diagrams. In these diagrams the solid line represents the non-critical interface and the dashed line the diffuse near-critical interface. Note that the asymptotic state is nonwet, with asymptotic contact angle $\hat\beta \approx 0.61$ rad in accord with the exact results derived in \cite{IK}. }
\label{fig:onequarter}
\end{figure}

For $t=0.25$ Fig.\ref{fig:onequarter} displays the interrelationship of the interfacial tensions $\sigma_{\alpha\beta}$ and $\sigma_{\beta\gamma}$. The generic law \eqref{generic} holds, and it also holds when $\sigma_{\alpha\beta}$ is replaced by $\sigma_{\alpha\gamma}$. Note that all states between the $\alpha\beta$ CEP and the $\beta\gamma$ CEP are nonwet, since \eqref{nonwet} holds throughout. 

\begin{figure}[h!]
\centering
\includegraphics[width=0.9\linewidth]{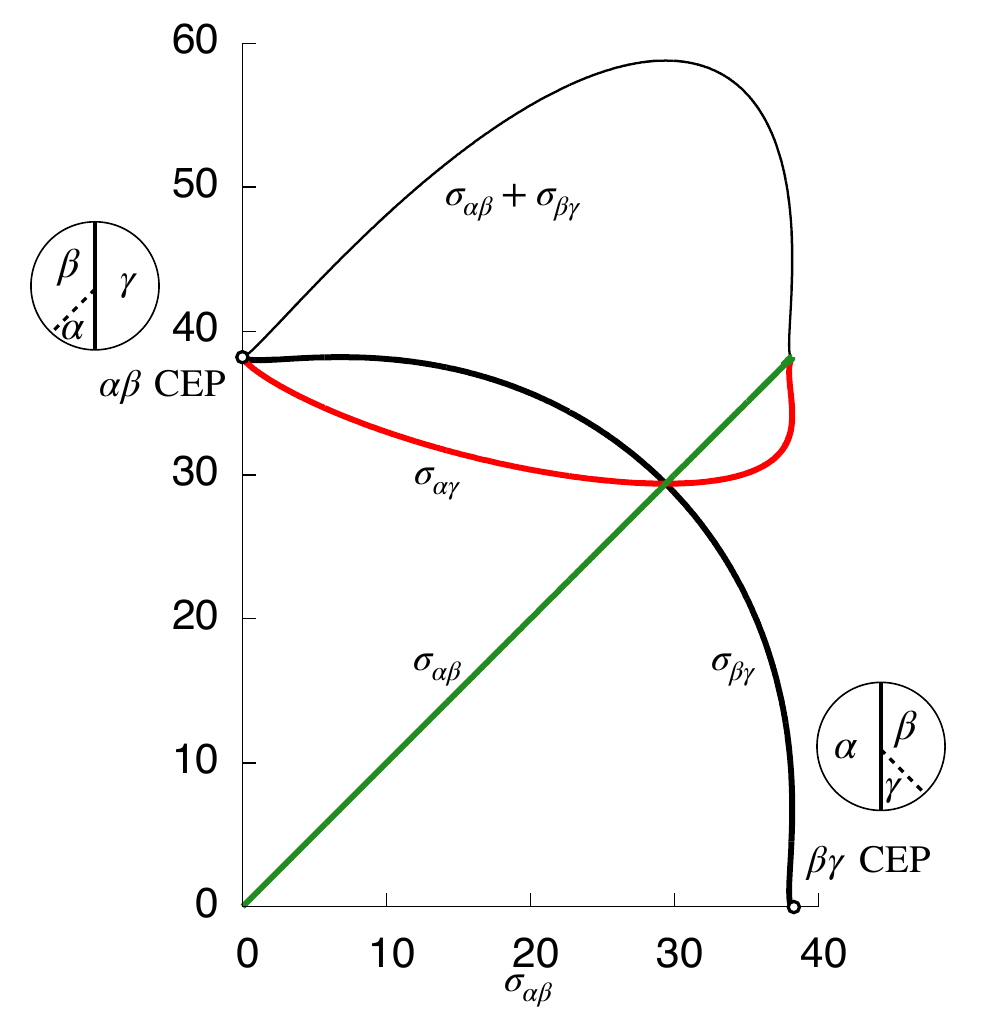}
\caption{Interfacial tensions $\sigma_{\alpha\beta}$, $\sigma_{\beta\gamma}$ and $\sigma_{\alpha\gamma}$ of the  interfaces $\alpha\beta$, $\beta\gamma$ and $\alpha\gamma$, respectively, versus $\sigma_{\alpha\beta}$ for a sweep (varying $s$) between the two CEPs at $t = 1$. Note that since $\sigma_{\alpha\gamma} < \sigma_{\alpha\beta} +\sigma_{\beta\gamma}$ only nonwetting states are encountered and CPW is absent. The CEPs lie in the nonwetting gap. Note that various interfacial tensions have become multi-valued near the $\beta\gamma$ CEP, when plotted as a function of $\sigma_{\alpha\beta}$, while they are single-valued functions of $s$ (not shown here).  The point where all three interfacial tensions intersect is the fully symmetric configuration with all three dihedral angles equal to $2\pi/3$. Note that the plot of $\sigma_{\beta\gamma}$ is mostly concave towards the origin, except very close to the CEPs where, once again, $\sigma_{\beta\gamma}$ approaches the axes tangentially (not visible on this scale), and features the exponent 3/2 in accord with \eqref{generic}.  The three-fluid interface configurations asymptotically close to the CEPs, on the edge of three-phase coexistence, are drawn in the diagrams. In these diagrams the solid line represents the non-critical interface and the dashed line the diffuse near-critical interface. Note that the asymptotic state is nonwet, with asymptotic contact angle $\hat\beta \approx 2.54 $ rad in accord with the exact results derived in \cite{IK}.}
\label{fig:one}
\end{figure}

For $t=1$ Fig.\ref{fig:one} likewise displays $\sigma_{\alpha\beta}$ and $\sigma_{\beta\gamma}$. The generic law \eqref{generic} holds, and it also holds when $\sigma_{\alpha\beta}$ is replaced by $\sigma_{\alpha\gamma}$. The nonwetting condition \eqref{nonwet} is satisfied along the entire sweep between the $\alpha\beta$ CEP and the $\beta\gamma$ CEP. Note that the plot of $\sigma_{\beta\gamma}$ is now mostly concave towards the origin instead of convex as in Fig.\ref{fig:onetenth}.

An interesting exception is found at the CEP that coincides with the endpoint of a neutral line (on which two interfacial tensions are equal). There are two such cases, related by symmetry, both occurring for $t=1/2$. They represent a surprising critical exponent anomaly, which was succinctly announced in \cite{KI}. The next subsection is devoted to its analysis.

\subsection{Critical exponent anomaly at the neutral point in the nonwetting gap}
\begin{figure}[h!]
\centering
\includegraphics[width=0.9\linewidth]{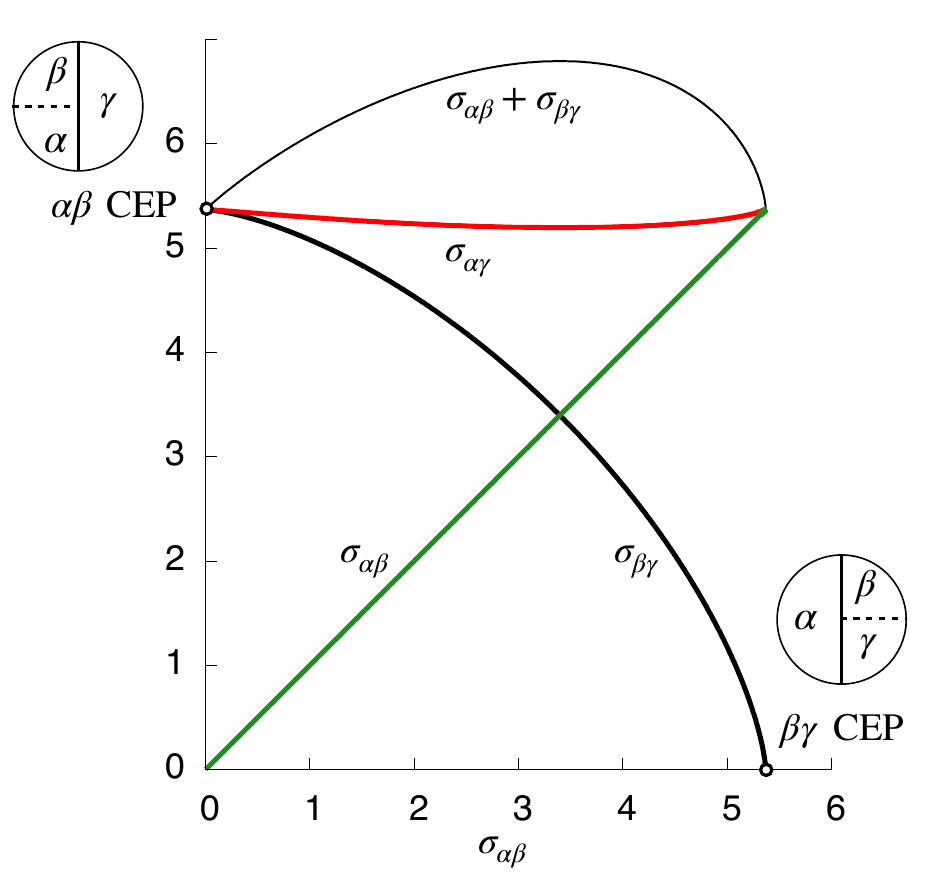}
\caption{Interfacial tensions $\sigma_{\alpha\beta}$, $\sigma_{\beta\gamma}$ and $\sigma_{\alpha\gamma}$ of the interfaces $\alpha\beta$, $\beta\gamma$ and $\alpha\gamma$, respectively, versus $\sigma_{\alpha\beta}$ for a sweep (varying $s$) between the two CEPs at $t = 1/2$. Here neutral lines meet the CEPs. \textcolor{black}{Note that the sweep is not along a neutral line.} Since $\sigma_{\alpha\gamma} < \sigma_{\alpha\beta} +\sigma_{\beta\gamma}$ only nonwetting states are encountered and CPW is absent. The CEPs lie in the nonwetting gap. Note that the plot of $\sigma_{\beta\gamma}$ is concave towards the origin. Asymptotically close to the CEPs, $\sigma_{\beta\gamma}$ approaches the axes orthogonally, and features the anomalous exponent 3/4 in accord with \eqref{anomaly1} and \eqref{anomaly2}.  The three-fluid interface configurations asymptotically close to the CEPs, on the edge of three-phase coexistence, are drawn in the diagrams. In these diagrams the solid line represents the non-critical interface and the dashed line the diffuse near-critical interface. Note that the asymptotic state is nonwet, with asymptotic contact angle $\hat\beta = \pi/2$. }
\label{fig:onehalf}
\end{figure}

When the neutral point, with $\sigma_{\alpha\beta} = \sigma_{\alpha\gamma}$, on the $\beta\gamma$ CEP line, at $t=1/2$, is approached from the three-phase region (see Fig.2 in \cite{IK}), \textcolor{black}{along any direction in the $(s,t)$-plane, } the generic law \eqref{generic} breaks down, and instead one finds, for $s>0$,
\begin{equation}
\label{anomaly1}
    \sigma_{\beta\gamma} \propto | \sigma_{\alpha\beta}-\sigma_{\alpha,\beta\gamma}|^{3/4},
\end{equation}
and, symmetrically, for $s<0$, interchanging $\alpha$ and $\gamma$,
\begin{equation}
\label{anomaly2}
    \sigma_{\alpha\beta} \propto  |\sigma_{\beta\gamma}-\sigma_{\gamma,\alpha\beta}|^{3/4}.
\end{equation}
This anomalous law is valid for an arbitrary trajectory in the $(s,t)$-plane, \textcolor{black}{with $\sigma_{\alpha\beta} \neq \sigma_{\alpha\gamma}$}, terminating on the CEPs at $t=1/2$. \textcolor{black}{The law is also valid for the special trajectory that follows the neutral line.} Fig.\ref{fig:onehalf} presents the interfacial tensions for a sweep at constant $t=1/2$. 

To understand the anomaly, a series expansion in $\epsilon$ is performed to order $\epsilon^2$, according to \eqref{expand}. For solutions $\rho_1^{\alpha} \leq \rho_1^{\beta} \leq \rho_1^{\gamma}$ of \eqref{three}, the expansions read, assuming $s>0$,
\begin{eqnarray}
\label{rhoa}
t^{-1/2}\rho_1^{\alpha} &=& -2 + \frac{2\epsilon}{9}  + \frac{8\epsilon^2}{243}  + {\cal O}(\epsilon^{5/2}) \\ \label{rhob}
t^{-1/2}\rho_1^{\beta} &=& 1 - \sqrt{\frac{2\epsilon}{3}} - \frac{\epsilon}{9}  \nonumber \\&-& \frac{5\epsilon^{3/2}}{54\sqrt{6}} - \frac{4\epsilon^2}{243}  + {\cal O}(\epsilon^{5/2})\\ \label{rhoc}
t^{-1/2}\rho_1^{\gamma} &=& 1 + \sqrt{\frac{2\epsilon}{3}}  - \frac{\epsilon}{9}  \nonumber \\
&+& \frac{5\epsilon^{3/2}}{54\sqrt{6}} - \frac{4\epsilon^2}{243}  + {\cal O}(\epsilon^{5/2}).
\end{eqnarray}

For the interfacial tension deviation from its critical-endpoint value the following expansion is obtained, with $\nu = \beta$ or $\gamma$,
\begin{eqnarray}
\label{expsigma}
    &&\sigma_{\alpha\nu}-\sigma_{\alpha,\beta\gamma} = -6 \sqrt{2}\, t^2(1-2t)^2 \epsilon \nonumber \\&\mp& \frac{8\,t^2}{3\sqrt{3}(1+t)} (1-2t)(-1 + 31 t - 4 t^2) \epsilon^{3/2} \nonumber \\&+& \frac{\sqrt{2}\,t^2 }{3(1+t)^{2} } (1+46t-435t^2+436t^3+52t^4) \epsilon^2 \nonumber \\&+& {\cal O}(\epsilon^{5/2}),
\end{eqnarray}
where the upper sign of the contribution of order $\epsilon^{3/2}$ holds for $\nu = \beta$ and the lower sign for $\nu = \gamma$. One may rewrite this result in a more insightful form, with $\nu = \beta$ or $\gamma$,
\begin{eqnarray}
\label{expsigma1}
    &&\sigma_{\alpha\nu}-\sigma_{\alpha,\beta\gamma} = -
    \frac{\sqrt{2}}{4}\, [(\vec{\rho}_{\alpha}-\vec{\rho}_{\beta\gamma})\cdot (\vec{\rho}_{\beta}-\vec{\rho}_{\gamma})]^2  \nonumber \\&\mp& \frac{1}{2}\sigma_{\beta\gamma} \,\cos \hat\beta + {\cal O}(\epsilon^{2}).
\end{eqnarray}
The first term in the r.h.s. of \eqref{expsigma1} is derived as follows. The dot denotes the scalar product of vectors in the $(\rho_1,\rho_2$)-plane. The bulk phase vector associated with  $\beta\gamma$ criticality is $\vec{\rho}_{\beta\gamma} = (t^{1/2},-t)$. Furthermore, using \eqref{rhoa}-\eqref{rhoc}, 
\begin{eqnarray}
    &\vec{\rho}_{\alpha}-\vec{\rho}_{\beta\gamma} = ((-3+\frac{2}{9}\epsilon)t^{1/2},-(3-\frac{8}{9}\epsilon)t) + {\cal O}(\epsilon^{2}),\\
    &\vec{\rho}_{\beta}-\vec{\rho}_{\gamma} = (-2\sqrt{\frac{2}{3}}\epsilon^{1/2}  t^{1/2},4\sqrt{\frac{2}{3}}\epsilon^{1/2}t)+ {\cal O}(\epsilon^{3/2})
\end{eqnarray}
and the scalar product is given by
\begin{equation}
    (\vec{\rho}_{\alpha}-\vec{\rho}_{\beta\gamma})\cdot (\vec{\rho}_{\beta}-\vec{\rho}_{\gamma}) =  6 \sqrt{\frac{2}{3}} \,t(1-2t)\,\epsilon^{1/2} + {\cal O}(\epsilon^{3/2})
\end{equation}
The next term in \eqref{expsigma1} features the asymptotic contact angle $\hat \beta $ between the $\alpha\beta$ and the $\beta\gamma$ interfaces infinitesimally close to the CEP line for $\beta\gamma$ criticality. The analytic expression for $\cos \hat \beta $ is given in \cite{IK}.  

The anomaly \eqref{anomaly1} can now be explained using geometrical arguments.
The first term on the r.h.s. of \eqref{expsigma1} pertains to the geometry of the three-phase triangle in the $(\rho_1,\rho_2$)-plane, in the vicinity of a CEP. It is zero when the triangle is isosceles, since the vector connecting $\alpha$ with the $\beta\gamma$ CEP is then perpendicular to the vector connecting the phase points $\beta$ and $\gamma$ that bifurcate from $\beta\gamma$. It is negative otherwise and its contribution is of first order in the thermodynamic field deviation $\epsilon$ away from the CEP. This signifies that in linear response, the interfacial tension deviation $\sigma_{\alpha\nu}-\sigma_{\alpha,\beta\gamma}$ is zero at $t=1/2$ where the line of CEPs intersects the neutral line. We conclude that a geometrical argument can explain the vanishing of the coefficient of $\epsilon$ for $t=1/2$ in the first term on the r.h.s. of \eqref{expsigma}.

The second term on the r.h.s. of \eqref{expsigma1} pertains to the geometry of the Neumann triangle \cite{RW}, which represents the triangle inequalities, such as \eqref{nonwet}, among the three interfacial tensions. Its edges are proportional to the interfacial tensions and its shortest edge is proportional to the critical interfacial tension $\sigma_{\beta\gamma}$.  In the vicinity of a CEP, where  $\sigma_{\beta\gamma}$ is of order $\epsilon^{3/2}$, the term under consideration vanishes when the asymptotic contact angle takes the value $\hat\beta = \pi/2$. Infinitesimally close to the CEP line, this happens when the Neumann triangle is isosceles, implying the symmetry $\sigma_{\alpha\beta} = \sigma_{\alpha\gamma}$ that characterises the neutral line. Again, the intersection of the CEP line with the neutral line singles out the value $t=1/2$, for which the coefficient of $\epsilon^{3/2}$ vanishes in the second term on the r.h.s. of \eqref{expsigma}. 

In sum, one is left with an interfacial tension deviation of order $\epsilon^2$ in place of $\epsilon$, with $\nu = \beta$ or $\gamma$,
\begin{equation}
\label{expsigmaneutral}
    \sigma_{\alpha\nu}-\sigma_{\alpha,\beta\gamma} = - \sqrt{2}\,  \epsilon^2 + {\cal O}(\epsilon^{5/2}), \,\, \mbox{for} \,\, t=\frac{1}{2}.
\end{equation}
On the other hand, the critical interfacial tension close to the CEP has, from \eqref{criticalinterface}, the following expansion,
\begin{equation}
\label{expsigmacritneutral}
    \sigma_{\beta\gamma} = 2 \sqrt{6} \, \epsilon^{3/2} + {\cal O}(\epsilon^{5/2}), \,\, \mbox{for} \,\, t=\frac{1}{2}.
\end{equation}
We conclude that the anomaly takes the following explicit form, near the $\beta\gamma$ CEP at $t=1/2$, with $\nu = \beta$ or $\gamma$,
\begin{equation}
\label{anomalyexplicit}
    \sigma_{\beta\gamma} =  2^{9/8}\sqrt{3} \,(\sigma_{\alpha,\beta\gamma}-\sigma_{\alpha\nu})^{3/4}.
\end{equation}

\section{Conclusion and an experimental perspective}
In fluid three-phase equilibria (with phases $\alpha,\beta,\gamma$) fluid-fluid interfacial tensions $\sigma$ have been studied in the vicinity of lines of critical endpoints that meet at a tricritical point in a mean-field density functional theory. Special attention has been paid to the behavior of $\sigma_{\alpha\gamma}$ and $\sigma_{\beta\gamma}$
as functions of $\sigma_{\alpha\beta}$ (at fixed temperature) in the three-phase coexistence region between the $\alpha\beta$ and $ \beta\gamma$ critical endpoints. The generic and expected law describing the vanishing of the critical interfacial tension as a function of the difference between the noncritical interfacial tension and its critical-endpoint value, has, with one notable exception, been verified to hold with the (mean-field) critical exponent 3/2. 

The exponent 3/2 has been confirmed for the segment of the CEP for which CPW occurs, as well as for the most part of the segment of the CEP for which CPW is absent and a nonwetting gap exists. However, an intriguing exception is found for one $\beta\gamma$ CEP in the nonwetting gap where the neutral line $\sigma_{\alpha\beta} = \sigma_{\alpha\gamma}$ meets the CEP line. When that CEP is approached \textcolor{black}{along any path, which need not coincide with the neutral line,} the mean-field exponent takes the value 3/4. This value is obtained exactly from analytic calculation and is explained, using geometrical arguments, to be the consequence of the simultaneous vanishing of two coefficients in the series expansion of the noncritical interfacial tension about the CEP. One coefficient vanishes because the three-phase triangle in the density plane becomes isosceles asymptotically close to the CEP and the other vanishes because the Neumann triangle becomes asymptotically isosceles around its shortest edge $\sigma_{\beta\gamma}$. 

Concerning the experimental relevance of the theoretical results described here, a number of remarks are in order. The Kahlweit-Busse and Aratono-Kahlweit experiments on phase-segregated oil, water and non-ionic amphiphile mixtures stated that, for a specific choice of mixture, no nonwetting to wetting transition was detected upon approaching either one of the two CEPs \cite{KahlweitBusse,AratonoKahlweit}. Indeed, near both CEPs for H$_2$O-$n$-octane-C$_8$E$_4$, and near the lower CEP for H$_2$O-$n$-octane-C$_6$E$_3$, detailed data suggested that the asymptotic contact angles are close to $\pi/2$ \cite{AratonoKahlweit}, instead of $0$ or $\pi$ expected for the CPW scenario. In contrast, for H$_2$O-$n$-octane-C$_5$E$_2$, their detailed data clearly display wetting transitions well before the CEPs are reached, illustrating CPW. 

Soon after these findings, however, an Erratum appeared in which Aratono and Kahlweit withdrew their statement concerning the possible persistence of nonwetting for H$_2$O-$n$-octane-C$_8$E$_4$ and H$_2$O-$n$-octane-C$_6$E$_3$, subsequent to finding a systematic error in the experiment. They announced revised experiments and added that they ``regret any irritation that our erroneous statement may have caused among theorists" \cite{Erratum}. Results from the later experiments on the same mixtures H$_2$O-$n$-octane-C$_8$E$_4$ and H$_2$O-$n$-octane-C$_6$E$_3$ indicated that the contact angles approaching the CEPs appear to be vanishing, which could still be consistent with a wetting transition, but perhaps too close to the CEP to be detectable \cite{KSB}.  

Clearly, experiments on these mixtures, or variants thereof, merit revisiting in the light of the possible consistency between experiment and the theoretical prediction of a nonwetting gap from exact solution of a DFT \cite{IK,Parry3}. The prediction of an anomalous critical exponent in the present paper, for the special case that the asymptotic non-wetting contact angle near the CEP is $\pi/2$, provides a new incentive for further experimental study. 

Acknowledgement. The authors dedicate this paper to the 
memory of their mentor Benjamin Widom. K.K. thanks KAKENHI (Grant Numbers 20H02696 and 25K00969). The authors thank Hiroki Matsubara for a discussion of the status of the cited experimental papers.

{}
\end{document}